\documentclass[prb,aps,twocolumn,amsmath,amssymb,floatfix,superscriptaddress]{revtex4}
\DeclareMathAlphabet{\mathpzc}{OT1}{pzc}{m}{it}
\usepackage[dvips]{graphics}
\usepackage{color}
\usepackage{enumitem}
\usepackage{soul}
\usepackage{mathrsfs}
\usepackage[colorlinks=true, citecolor=blue, urlcolor=blue]{hyperref}

\definecolor{dred}{rgb}{0.75,0,0}

\date{\today}

\begin{document}

\title{Antiferromagnetic helix as an efficient spin polarizer: Interplay between electric field and higher ordered hopping}

\author{Debjani Das Gupta}

\affiliation{Physics and Applied Mathematics Unit, Indian Statistical
Institute, 203 Barrackpore Trunk Road, Kolkata-700 108, India}

\author{Santanu K. Maiti}

\email{santanu.maiti@isical.ac.in}

\affiliation{Physics and Applied Mathematics Unit, Indian Statistical
Institute, 203 Barrackpore Trunk Road, Kolkata-700 108, India}

\begin{abstract}

We report spin filtration operation considering an antiferromagnetic helix system, possessing zero net magnetization. Common wisdom
suggests that for such a system, a spin-polarized current is no longer available from a beam of unpolarized electrons. But, once we apply
an electric field perpendicular to the helix axis, a large separation between up and down spin energy channels takes place which yields
a high degree of spin polarization. Such a prescription has not been reported so far to the best of our concern. Employing a tight-binding
framework to illustrate the antiferromagnetic helix, we compute spin filtration efficiency by determining spin selective currents using
Landauer-B\"{u}ttiker formalism. Geometrical conformation plays an important role in spin channel separation, and here we critically
investigate the effects of short-range and long-range hoppings of electrons in presence of the electric field. We find that the filtration
performance gets improved with increasing the range of hopping of electrons. Moreover, the phase of spin polarization can be altered
selectively by changing the strength and direction of the electric field, and also by regulating the physical parameters that describe the 
antiferromagnetic helix. Finally, we explore the specific role of dephasing, to make the system more realistic and to make the present
communication a self-contained one. Our analysis may provide a new route of getting conformation-dependent spin polarization 
possessing longer range hopping of electrons, and can be generalized further to different kinds of other fascinating antiferromagnetic 
systems.

\end{abstract}

\maketitle

\section{Introduction}

After the discovery of the giant magnetoresistance (GMR) effect~\cite{gmr1,gmr2,prinz}, {\em spintronics} becomes a new field of 
research in the discipline of condensed matter physics where the spin degree of freedom of an electron is explored along with its 
charge~\cite{wolf,spin1,spin2,spin3}. In conventional electronic devices, Joule heating is an inevitable effect due to the flow of electrons
which causes a sufficient power loss. But, if we utilize electron spin instead of the charge, then power will be consumed and at the same
time operation will be much faster~\cite{spin1,spin2}. Two pivotal features for the consideration of spin-based electronic devices rather
than conventional charged-based ones are: (i) saving more power and information can be transferred at a much faster rate which 
undoubtedly reduces cost price by a significant amount and (ii) size of the devices becomes too small so that a large number of 
functional elements can be integrated into a small dimension~\cite{wolf,spin1}. For instance, the hard disk drive made in 1957 was 
able to store data only up to $3.75$ megabytes, and it occupied a volume of $68$ cubic feet. Whereas, a recent hard disk drive 
possessing a volume of the order of $2.1$ cubic inches can even store data up to several terabytes~\cite{hd}. Using the GMR phenomenon, 
not only in storing devices but significant development has been made in different other technologies involving electron 
spin~\cite{spin1,spin2,spin3}.

One of the most fundamental issues in spintronics is to find an efficient route for the separation of two spins, or more precisely we
can say, the generation of polarized spin current from a completely unpolarized electron beam. Several propositions have already been
made along this line~\cite{curr1,curr2,fm1}. The most common practice is to use ferromagnetic materials, though there are several
unavoidable limitations~\cite{fm2}. For instance, a large resistivity mismatch occurs across a junction formed by ferromagnetic and
non-magnetic materials, which hinders the proper injection of electrons into the system~\cite{fm2,fm3}. The other crucial limitation arises
when we think about the tuning spin selective junction currents. Usually, this is done by means of an external magnetic field, but for a
quantum regime, it is very hard to confine a strong magnetic field, and the problem still persists even today. Over the last few years,
the use of ferromagnetic materials significantly gets suppressed when spin-orbit (SO) coupled systems came into the picture. Two different
kinds of SO interactions are taken into account in solid-state materials, one is known as Rashba~\cite{rashba} and the other one is
referred to as Dresselhaus SO interaction~\cite{dres,soi}. The latter type appears due to the breaking of bulk inversion symmetry of
a system, whereas the previous one arises due to the breaking of the symmetry in confining potential. Among these two, the Rashba
strength can be tuned externally by suitable setups~\cite{gate1,gate2}, and therefore, the Rashba SO coupled systems draw significant
attention than the Dresselhaus ones in the field of spintronics. Different kinds of systems starting from tailor-made geometries, organic and 
inorganic molecules have been considered as functional elements in two-terminal as well as multi-terminal setups, and many interesting
features have been explored~\cite{multi1,multi2,multi3}. But, in most of these cases, especially in molecular systems, the major concern
is that the SO coupling strength is too weak compared to the electronic hopping strength, almost an order of magnitude
smaller~\cite{strength}. Moreover, the variation of the SO coupling strength is also quite limited by external means. Because of these
facts, a high degree of spin polarization and its possible tuning in a wide range are quite difficult to achieve in SO coupled systems,
though there are of course many other advantages that make these systems promising functional elements in spintronics.

To avoid all these issues, modern machinery has concentrated on antiferromagnetic (AF) materials that possess an alternate type of magnetic
ordering and have {\em zero net magnetization}~\cite{afm1,afm2,afm3}. Several key prospects of using an AF system as a spin-polarized 
functional element are there. For instance, these materials are insensitive to external magnetic fields, and they are much faster and 
can be operated up to a high-frequency range ($\sim$THz) than the traditional ferromagnetic systems~\cite{thz}. Moreover, due to the
absence of any stray fields, a large number of closely packed functional elements can be accommodated in a small region which leads to
several important advantages in designing efficient electronic devices based on spin-based transport phenomena. Nowadays,
antiferromagnetic spintronics evolves as a cutting-edge research field, and, may lead to new prospects in the magnetic
community~\cite{afm4,afm5,afm6,afm7}.

In the present work, we propose a new prescription for efficient spin filtration considering an antiferromagnetic system, which we refer to 
as an ``antiferromagnetic helix" (AFH). The role of chirality on spin filtration first came into realization based on the experimental work
of G\"{o}hler {\em et al.}~\cite {dna} where they have shown that almost $60\%$ spin polarization can be achieved through a self-assembled
monolayers of double-stranded DNA molecules deposited on a gold substrate. They have described this effect as {\em chiral induced spin
selectivity} (CISS). After this realization, several experimental and theoretical research groups have paid significant attention to 
this CISS effect, considering different kinds of molecular as well as artificially designed systems, possessing helical 
geometry~\cite{prl,protein,bacteria,dna1,helix,aprotein,edge,sarkar,magfield}. But, to the best of our knowledge, no effort has been
made so far to address the phenomenon of spin filtration considering a `magnetic helix structure with vanishing net magnetization',
and this is precisely the fundamental motivation behind the present work. For our AFH, simulated by a tight-binding framework, magnetic
moments in alternate lattice sites are arranged in opposite directions, resulting in a zero net magnetization. In such a system, common
wisdom suggests that spin filtration is no longer possible. But, interestingly we find that, once we apply an electric field perpendicular
to the helix axis, a large separation between two spin channels takes place, which results in a high degree of spin polarization. The
central mechanism relies on the helicity and the applied electric field. In the absence of any of these two, helicity and electric field,
no such phenomenon is observed.

Determining spin-dependent transmission probabilities using the well-known Green's function formalism~\cite{gf1,gff1,gf2}, we compute spin
selective currents through the AFH following the Landauer-B\"{u}ttiker prescription~\cite{gf3,gf4}. From the currents, we evaluate the spin 
polarization coefficient. The geometrical conformation plays a significant role in spin filtration, and we investigate it by considering
both short- and long-range hopping cases. From our analysis, we find that the spin filtration efficiency gets enhanced with increasing
electron hopping among more lattice sites. The specific roles of all other physical quantities are thoroughly discussed which
lead to several interesting features. Finally, to make the quantum system more realistic and for the sake of completeness of our study,
we include the effects of electron dephasing~\cite{dephase1,dephase2,dephase3,dephase4,dephase5} on spin polarization. Our analysis may
provide some key inputs towards designing efficient spintronic devices considering different kinds of antiferromagnetic helical systems,
possessing longer range hopping of electrons.

The rest part of the work is arranged as follows. 
\begin{figure}[ht]
{\centering\resizebox*{8cm}{7cm}{\includegraphics{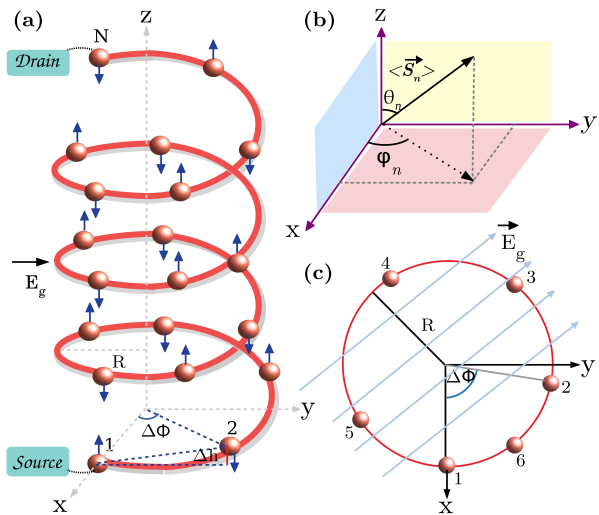}}\par}
\caption{(Color online). (a) A right-handed antiferromagnetic helix is clamped between two one-dimensional non-magnetic electrodes, 
namely, source and drain. $R$ being the radius of the helix, and $\Delta \phi$ and $\Delta h$ correspond to the twisting angle and 
stacking distance between the magnetic sites (colored balls), respectively. Successive magnetic moments, shown by the blue arrows, 
associated with the magnetic sites are arranged along $+Z$ and $-Z$ directions, resulting in a vanishing net magnetization. An electric 
field of strength $E_g$ is applied perpendicular to the helix axis, which plays a pivotal role in spin filtration. (b) Orientation of 
the net spin $\langle \boldsymbol{\vec{S}_n}\rangle$ in spherical polar coordinate system. (c) Projection of the right-handed helix 
on the $X$-$Y$ plane.}
\label{model}
\end{figure}
Section II includes the spin-polarized setup, tight-binding (TB) Hamiltonian of the junction, and the required theoretical prescriptions 
for the calculations. All the results are presented and thoroughly discussed in Sec. III. Finally, the essential findings are summarized 
in Sec. IV.

\section{Quantum system, TB Hamiltonian and theoretical formulation}

\subsection{Junction setup and the TB Hamiltonian}

Let us begin with the spin-polarized setup shown in Fig.~\ref{model}(a), where an antiferromagnetic (AF) system is coupled to the source 
and drain electrodes. The magnetic sites (filled colored balls) in the AF system are arranged in a helical pattern. Each of these sites
contains a finite magnetic moment, denoted by the blue arrow, associated with a net spin $\langle\vec{\boldsymbol S_n}\rangle$.
The successive magnetic moments are aligned in opposite directions ($\pm Z$), and therefore, the net magnetization of the helix becomes
zero. We refer to this system as an ``antiferromagnetic helix" (AFH), and here in our work, we will show how such an AFH acts for spin
filtration. The general orientation of any local spin $\langle\vec{\boldsymbol S_n}\rangle$, and hence the magnetic moment, can be
illustrated by the usual co-ordinate system as presented in Fig.~\ref{model}(b), where $\theta_n$ and $\varphi_n$ are the polar and
azimuthal angles, respectively.

Two essential physical parameters characterize the helical geometry~\cite{prl,protein} those are $\Delta \phi$ and $\Delta h$ (see 
Fig.~\ref{model}), where the first one represents the twisting angle and the other parameter denotes the stacking distance. Depending 
on $\Delta h$, we can have a helical system where lower or higher-order hopping of electrons becomes significant. When $\Delta h$ is 
very small i.e., atoms in the helix are densely packed, electrons can hop at multiple sites, which is referred to as `long-range 
hopping antiferromagnetic helix' (LRH AFH). On the other hand, when the atoms are less densely packed viz, $\Delta h$ is quite large, 
electrons can hop between a few neighboring magnetic sites. Such a system is called a `short-range hopping antiferromagnetic helix' 
(SRH AFH). In the present work, we consider these two different kinds of AF helices and investigate the results.

The antiferromagnetic helix is subjected to an electric field, having strength $E_g$, perpendicular to the helix axis. It plays a
central role in spin filtration, and it can be understood from our forthcoming discussion. With a suitable setup, one can tune 
the strength of this field as well as its direction.

In order to investigate spin dependent transport phenomena and to exhibit the spin filtration operation, the helical system is clamped
between source and drain electrodes. We describe this nanojuntion using the tight-binding framework~\cite{sk1,sk2,sk3,skm1}. The TB 
Hamiltonian of the full system is written as a sum
\begin{eqnarray}
\boldsymbol{\mathit{H}}=\boldsymbol{\mathit{H}}_{AFH} + 
\boldsymbol{\mathit{H}}_{S} + \boldsymbol{\mathit{H}}_{D} + \boldsymbol{\mathit{H}}_{tun}
\label{total}
\end{eqnarray}
where different sub-Hamiltonians in the right side of Eq.~\ref{total} are associated with different parts of the nanojunction and they 
are described as follows. 

The term $\boldsymbol{\mathit{H}}_{AFH}$ corresponds to the Hamiltonian of the antiferromagnetic helix. For an AFH, be it a short-range 
or a long-range one, the TB Hamiltonian is expressed as
\begin{eqnarray}
\boldsymbol{\mathit{H}}_{AFH} = \sum_n \boldsymbol{c}_n^{\dagger}
\left (\boldsymbol{\epsilon}_n -\boldsymbol{\vec{\mathpzc{h}}}_n\cdot
\boldsymbol {\vec{\sigma}}\right )
\boldsymbol{c}_n \nonumber\\
+ \sum_{n=1}^{N-1} \sum_{m=1}^{N-n}\left (\boldsymbol{c}_n^{\dagger} \boldsymbol{t}_m
\boldsymbol{c}_{n+m} + h.c. \right ).
\label{hamil}
\end{eqnarray}
where $\boldsymbol{c}_n^{\dagger} = \left(c_{n\uparrow}^{\dagger}\quad c_{n\downarrow}^{\dagger} \right)$. 
$c_{n\sigma}^{\dagger}$ $(c_{n\sigma})$ is the creation (annihilation) operator of an electron at site $n$ with spin 
$\sigma\left(= \uparrow,\downarrow\right)$. 
$\boldsymbol{\epsilon}_n-\vec{\boldsymbol{\mathpzc{h}}}_n\cdot\vec{\boldsymbol{\sigma}}$ is the effective site energy matrix
which looks like
\begin{equation}
\boldsymbol{\epsilon}_n-
\vec{\boldsymbol{\mathpzc{h}}}_n\cdot\vec{\boldsymbol{\sigma}}=
\begin{pmatrix}
\epsilon_n -\mathpzc{h}\cos\theta_n & -\mathpzc{h}\sin\theta_n e^{-i\varphi_n}\\
-\mathpzc{h}\sin\theta_n e^{i\varphi_n} & \epsilon_n +\mathpzc{h}\cos\theta_n
\end{pmatrix} 
\label{sitemat}
\end{equation}
where $\epsilon_n$ is the on-site energy in the absence of any kind of magnetic scattering.
$\boldsymbol{\vec{\mathpzc h}}_n = J\langle \boldsymbol{\vec{S}}_n \rangle$, called as the spin-flip scattering parameter where $J$ is
the coupling strength~\cite{strength} between the coupling of an itinerant electron with local magnetic moment, associated with the
average spin $\langle \boldsymbol{\vec{S}}_n \rangle$. $\vec{\boldsymbol{\sigma}}$ is the Pauli spin vector. Here we assume that
$\boldsymbol{\sigma}_z$ is diagonal. The term $\vec{\boldsymbol{\mathpzc{h}}}_n\cdot\vec{\boldsymbol{\sigma}}$ represents the 
spin-dependent scattering and it is widely used in literature~\cite{sk1,sk2,sk3,skm1}. The key point is that the strength `$J$' is 
reasonably large than the other spin-dependent scattering parameters viz, SO coupling, Zeeman splitting in presence of magnetic field, 
etc~\cite{strength}. Thus, there is a possibility of getting a high degree of spin filtration under suitable input condition(s), in 
the presence of a spin-moment scattering mechanism.

The second term of Eq.~\ref{hamil}, involving $\boldsymbol{t}_m$, is quite tricky, not like usual nearest-neighbor 
hopping case, and the summations over $n$ and $m$ need to take carefully. $\boldsymbol{t}_m$ is a $(2 \times 2)$ 
hopping matrix, and it becomes 
\begin{eqnarray}
\boldsymbol{t}_m =
\begin{pmatrix}
t_m & 0\\
0 & t_m
\end{pmatrix}
\end{eqnarray}
where $t_m$ represents the hopping between the sites $n$ and $(n+m)$. The hopping strength $t_m$
is written as~\cite{prl,protein}
\begin{equation}
t_m = t_1 \textrm{e}^{-(l_m-l_1)/l_c}
\label{hop}
\end{equation}
where $t_1$ is the nearest-neighbor hopping integral, $l_m$ is the distance of separation between the sites $n$ and 
$(n+m)$, $l_1$ is the distance among the nearest-neighbor sites and $l_c$ is the decay constant. In terms of the radius $R$ 
(see Fig.~\ref{model}(c) where the projection of the helix in the $X$-$Y$ plane is shown), twisting angle $\Delta \phi$ and the stacking 
distance $\Delta h$, $l_m$ gets the form~\cite{prl,protein,edge} 
\begin{equation}
l_m = \sqrt {{[2 R \sin (m \Delta \phi/2)}]^2 + (m \Delta h)^2}.
\label{lm}
\end{equation}

When the AFH is subjected to a transverse electric field, its site energies get modified. The effective site energy for any site $n$ 
becomes~\cite{aprotein,edge} 
\begin{equation}
\epsilon_n^{\rm eff} =
\epsilon_n +eV_g \cos(n\Delta \phi-\beta)
\label{eg}
\end{equation}
where $e$ is the electronic charge, and, $V_g$  ($=2E_g R$) is the gate voltage, responsible for the generation of the electric field. 
$\beta$ represents the angle between the incident electric field and the positive $X$-axis~\cite{prl,protein}.

The TB Hamiltonians of the side attached source (S) and drain (D) electrodes, $\boldsymbol{\mathit{H}}_{S}$ and 
$\boldsymbol{\mathit{H}}_{D}$ and their coupling with the AFH ($\boldsymbol{\mathit{H}}_{tun}$) look quite simple than what is 
described above for the antiferromagnetic helix. The electrodes are assumed to be perfect, one-dimensional, and non-magnetic in 
nature. They are expressed as, 
\begin{eqnarray}
\boldsymbol{\mathit{H}}_{S}=\boldsymbol{\mathit{H}}_{D}=
\sum_n \boldsymbol{a}_n^{\dagger}\boldsymbol{\epsilon}_0
\boldsymbol{a}_n
+ \sum_{n}\left (
\boldsymbol{a}_{n+1}^{\dagger} \boldsymbol{t}_0
\boldsymbol{a}_{n} + h.c. \right )
\label{lead}
\end{eqnarray}
where $\boldsymbol{\epsilon}_0 = {\rm diag} (\epsilon_0,\epsilon_0)$ and 
$\boldsymbol{t}_0={\rm diag}(t_0,t_0)$. 
$\epsilon_0$ and $t_0$ are the on-site energy and nearest-neighbor hopping integral, respectively. 
$\boldsymbol{a}_n^{\dagger} = \left(a_{n\uparrow}^{\dagger}\quad a_{n\downarrow}^{\dagger} \right)$, 
$a_{n\sigma}$'s are the usual fermionic operators in the electrodes. 

Finally, the tunneling Hamiltonian is expressed as 
\begin{eqnarray}
\boldsymbol{\mathit{H}}_{tun} = t_S \left(\boldsymbol{c}_1^{\dagger}\boldsymbol{a}_{-1} + h.c.\right) +
t_D \left(\boldsymbol{c}_N^{\dagger}
\boldsymbol{a}_{N+1} + h.c. \right )
\label{tunnel}
\end{eqnarray}
where $t_S$ and $t_D$ are the coupling strengths of the AFH with S and D, respectively. We refer to the lattice site of the source which 
is coupled to the helix as $-1$ and the site of the drain which is attached to the helix as $N+1$. The sites $1,2 \dots N$ ($N$ being
the total number of magnetic sites) are used for the AFH.

\subsection{Theoretical formulation}

In order to inspect spin-dependent transport phenomena and spin polarization coefficient, the first and foremost thing that we need to 
calculate is the two-terminal transmission probability. We compute it using the well known non-equilibrium Green's function 
formalism~\cite{gf1,gff1,gf2,gf3,gf4}. In terms of retarded and advanced Green's functions, $\boldsymbol{\mathit{G}}^r$ 
and $\boldsymbol{\mathit{G}}^a$, the spin-dependent transmission coefficient is obtained from the expression~\cite{gf1,gff1,gf4}
\begin{equation}
T_{\sigma \sigma^{\prime}}= \textrm{Tr} \left[\boldsymbol{\Gamma}_S \boldsymbol{\mathit{G}}^r \boldsymbol{\Gamma}_D
\boldsymbol{\mathit{G}}^a \right]
\label{trans}
\end{equation}
where, 
\begin{equation}
\boldsymbol{\mathit{G}}^r=\left(\boldsymbol{\mathit{G}}^a\right)^{\dagger}=\left[\boldsymbol{E}-
\boldsymbol{\mathit{H}}_{AFH}-
\boldsymbol{\Sigma}_S - \boldsymbol{\Sigma}_D\right]^{-1}.
\label{green}
\end{equation}
$\boldsymbol{\Sigma}_S$ and $\boldsymbol{\Sigma}_D$ are the self-energy matrices~\cite{gf1,gff1,gf2} which capture all the essential 
information of the electrodes and their coupling with the helix. $\boldsymbol{\Gamma}_S$ and $\boldsymbol{\Gamma}_D$ 
are the coupling matrices.

For an incoming electron with spin $\uparrow$, two things may occur. We can have a finite possibility of getting up spin 
electron as a up spin or it can be flipped. Similar options are also available for an injected down spin electron. 
Thus, considering pure $(T_{\sigma \sigma})$ and spin flip $(T_{\sigma \sigma^{\prime}})$ transmissions, we can write the net up 
and down spin transmission probabilities ($T_{\uparrow}$ and $T_{\downarrow}$) as 
\begin{subequations}
\begin{align}
T_{\uparrow} &=T_{\uparrow,\uparrow} + T_{\downarrow,\uparrow} \\
T_{\downarrow} &=T_{\downarrow,\downarrow} + T_{\uparrow,\downarrow}.
\end{align}
\end{subequations}

From the transmission coefficients $T_{\uparrow}$ and $T_{\downarrow}$, we evaluate up and down spin junction currents, using the Landauer 
B\"{u}ttiker prescription~\cite{gf1,gff1,gf2,gf3,gf4}. The spin-dependent current, when a finite bias $V$ is applied across the AFH, is 
expressed as 
\begin{equation}
I_{\sigma} = \frac{e}{h} \int T_{\sigma} \left(f_S-f_D \right) dE
\label{current}
\end{equation}
where, $f_S$ and $f_D$ are the Fermi functions, associated with S and D, respectively, and they are
\begin{equation}
f_{S(D)}=\frac{1}{1+\textrm{e}^{\left(E-\mu_{S(D)}\right)/k_B \mathsf{T}}}.
\label{fermi}
\end{equation}
Here $\mu_S$ and $\mu_D$ are the electro-chemical potentials of S and D, respectively, and $k_B \mathsf{T}$ is the thermal energy.

Determining $I_{\uparrow}$ and $I_{\downarrow}$, we evaluate spin filtration efficiency following the relation~\cite{dephase3} 
\begin{equation}
P = \frac{I_{\uparrow}-I_{\downarrow}} {I_{\uparrow}+I_{\uparrow}}\times 100\%.
\label{pol}
\end{equation}
When only up spin electrons propagate we get $P=100\%$, while for the situation where only down spin electrons get transferred through
the AFH, we get $P=-100\%$. For the situation where both up and down spin electrons propagate equally, no spin filtration occurs. We want 
to reach the limiting value where $P= 100\%$ or $-100\%$, which is usually very hard to achieve.

\vskip 0.2cm
\noindent
\underline{\em Inclusion of dephasing}: Dephasing is an important factor and in many cases, it cannot be avoided especially when we think
about the experimental realization of a theoretical proposal. There are different possible routes through which a system is disturbed 
by dephasing, and it is thus required to incorporate its effect in our analysis.
Several methodologies~\cite{dephase1,dephase2,dephase3,dephase4,dephase5} are available for the inclusion of dephasing and most of them 
are very complex. B\"{u}ttiker on the other hand predicted phenomenologically a very simple but elegant way to incorporate the dephasing 
effect into the system~\cite{dephase1,dephase2}, and here we use the same procedure. In this prescription it is assumed that each 
lattice site of the AFH is attached to a dephasing electrode, commonly referred to as B\"{u}ttiker probe. The key concept is that the 
dephasing electrodes will not drag or inject any finite number of electrons into the system i.e., the net current passing through such 
electrodes becomes exactly zero~\cite{dephase1,dephase2}. Electrons from the AFH enter into the dephasing electrodes, and after losing 
their phase memories, they eventually come back to the parent system. 

In order to achieve the zero current condition in different dephasing electrodes, we need to choose the voltages $V_n$ ($V_n$ being 
the voltage at $n$th dephasing electrode) in the appropriate way~\cite{aprotein,gff1}. The $V_n$'s are determined following the 
Landauer-B\"{u}ttiker current expression~\cite{gf1,gff1} associated with each dephasing electrode, and evaluating the bias drop at 
different lattice sites of the helix. It is crucial to point out that, the evaluation of this bias drop is quite complicated
as it is a non-linear problem. The prescription can be simplified to some extent by considering a linear profile along the helix, 
which is most commonly used in literature, and here in our present work we also follow it. Suppose a finite voltage $V_0$ is applied 
between the real electrodes $S$ and $D$, and (say) $V_S=V_0$ and $V_D=0$, without loss of any generality. Then, the voltages $V_n$
at different lattice sites can be calculated without much difficulty, as we assume the linear drop, and adjusting these voltages 
across the dephasing electrodes, the zero-current condition is established (a more detailed discussion about it is available 
in~\cite{aprotein,gff1}).

In presence of the B\"{u}tiker probes, the transmission probability of getting electrons at the drain electrode (D) is 
modified~\cite{aprotein,dephase5}, and it becomes
\begin{equation}
T_{\sigma \sigma^{\prime}}^{eff}(V) = \sum_{\alpha=S,n} T_{\sigma \sigma^{\prime}}^{\alpha D}(V) \frac{V_{\alpha}}{V_0}.
\label{dephase}
\end{equation}
The transmission probabilities are now voltage dependent, and thus, special care has to be taken to calculate these 
quantities~\cite{vt1,vt2}. 
Here $T_{\sigma \sigma'}^{S D} (V)$ and $T_{\sigma \sigma'}^{n D} (V)$ denote the transmission probabilities from the source 
electrode (S) and from the $n$-th dephasing electrode to the drain end, respectively. The dephasing electrodes are connected 
at all the lattice sites of the antiferromagnetic helix, apart from the sites $1$ and $N$ where the real electrodes (S, D) are 
attached. In our formulation, the coupling strength between the AFH and the dephasing electrode is mentioned by the parameter 
$\eta$, and it describes the dephasing strength.

To compute $T_{\sigma \sigma^{\prime}}^{eff}(V)$, an important step must be performed which is as follows. For a biased system,
since the scattering states become the eigenstates of the biased Hamiltonian, the site energy $\epsilon_0$ needs to be shifted
by $V_0$ in the source electrode, and by $V_n$ in the $n$-th B\"{u}ttiker probe~\cite{vt1,vt2}. Using Eq.~\ref{dephase}, we get 
the effective up and down spin transmission probabilities, at different voltages, from the relations
\begin{subequations}
\begin{align}
T_{\uparrow}^{eff}(V) &=T_{\uparrow,\uparrow}^{eff}(V) + T_{\downarrow,\uparrow}^{eff}(V) \\
T_{\downarrow}^{eff}(V) &=T_{\downarrow,\downarrow}^{eff}(V) + T_{\uparrow,\downarrow}^{eff}(V).
\end{align}
\end{subequations}
The effective spin-dependent current in presence of dephasing can thus be obtained through the expression
\begin{equation}
I_{\sigma}(V) = \frac{e}{h} \int T_{\sigma}^{eff}(V) \left(f_S-f_D \right) dE.
\label{currdph}
\end{equation}
With the effective spin-dependent currents, the same definition is followed as mentioned in Eq.~\ref{pol}, to compute spin polarization 
coefficient $P$ in the presence of dephasing.

In the extreme low biased condition, the above current equation (Eq.~\ref{currdph}) for any $q$-th electrode, be it real or virtual, 
boils down to~\cite{aprotein}
\begin{equation}
I_{\sigma}^q(V) = \frac{e^2}{h} \sum_{\alpha} T_{\sigma}^{\alpha q} (V_{\alpha}-V_q)
\label{lb}
\end{equation}
where the voltages $V_{\alpha}$ and $V_q$ can be derived from the prescription given above. In this limiting condition, the current is
`linearly' proportional to the voltage. On the other hand, in the limit of high bias, Eq.~\ref{currdph} cannot be simplified in the 
linear form like what is given in Eq.~\ref{lb}, and we get the non-linear behavior. For the sake of completeness of our analysis, 
we discuss the accuracy of the above prescription in the appropriate sub-section.

\section{Numerical Results and Discussion}

Now we present our results and investigate the specific role of the external electric field on spin filtration, under different input 
conditions. Both the short-range and long-range antiferromagnetic helices are taken into account. For these two types of AFHs, we choose
the geometrical parameters $R$, $\Delta h$, and $\Delta \phi$ as given in Table~\ref{table1}. 
These parameter values are analogous to 
the real helical systems like single-stranded DNA and protein molecules, and they are the most suitable examples where respectively the 
short-range and long-range hopping models are taken into account~\cite{prl,protein}. A large amount of investigation has already been 
done in the literature considering this particular set of parameter values in different contemporary works, and accordingly, here also
we select these typical values. Other sets of parameter values represent the 
SRH and LRH of electrons can also be considered, and all the physical pictures studied here will remain unaltered.

The other physical parameters those are common throughout the analysis are as follows. In the absence of electric field, the on-site 
\begin{table}[ht]
\begin{center}
\caption{Geometrical parameters describing the SRH and LRH AFHs.}
\vskip 0.2cm
\begin{tabular}{|c||c|c|c|c|}
\hline
System & $R$ (nm) & $\Delta h$ (nm) & $\Delta \phi$ (rad) & $l_c$ (nm) \\ \hline
SRH AFH & $0.7$ & $0.34$ & $\pi/5$ & $0.09$ \\ \hline
LRH AFH & $0.25$ & $0.10$ & $5\pi/9$ & $0.09$ \\ \hline
\end{tabular}
\label{table1}
\end{center}
\end{table}
energies ($\epsilon_n$) in the AFH are set to zero, and we fix the NNH strength $t_1 = 1\,$eV. The spin dependent scattering parameters 
$\mathpzc{h}$ is set at $1\,$eV. As already mentioned, the successive magnetic moments in the helix system are aligned 
in opposite directions ($\pm Z$) (see the schematic diagram given in Fig.~\ref{model}(a)). We set $\theta_n=0$ for all odd $n$ sites, and 
$\theta_n=\pi$ for all the even $n$ sites. The azimuthal angle $\varphi_n$ is fixed to zero for all $n$. In the side attached electrodes, 
we choose $\epsilon_0=0$, $t_0 = 2\,$eV. The coupling parameters $t_S$ and $t_D$ are set at $1\,$eV. All the other energies are also 
measured in units of electron-volt (eV). Unless specified, the results are worked out considering a right-handed antiferromagnetic helix 
with $\beta=0$ and in the absence of dephasing. We set the system temperature at $100\,$K, throughout the discussion.

\subsection{Spin dependent transmission probabilities and spin polarization coefficient}

\begin{figure}[ht]
{\centering\resizebox*{8cm}{6cm}{\includegraphics{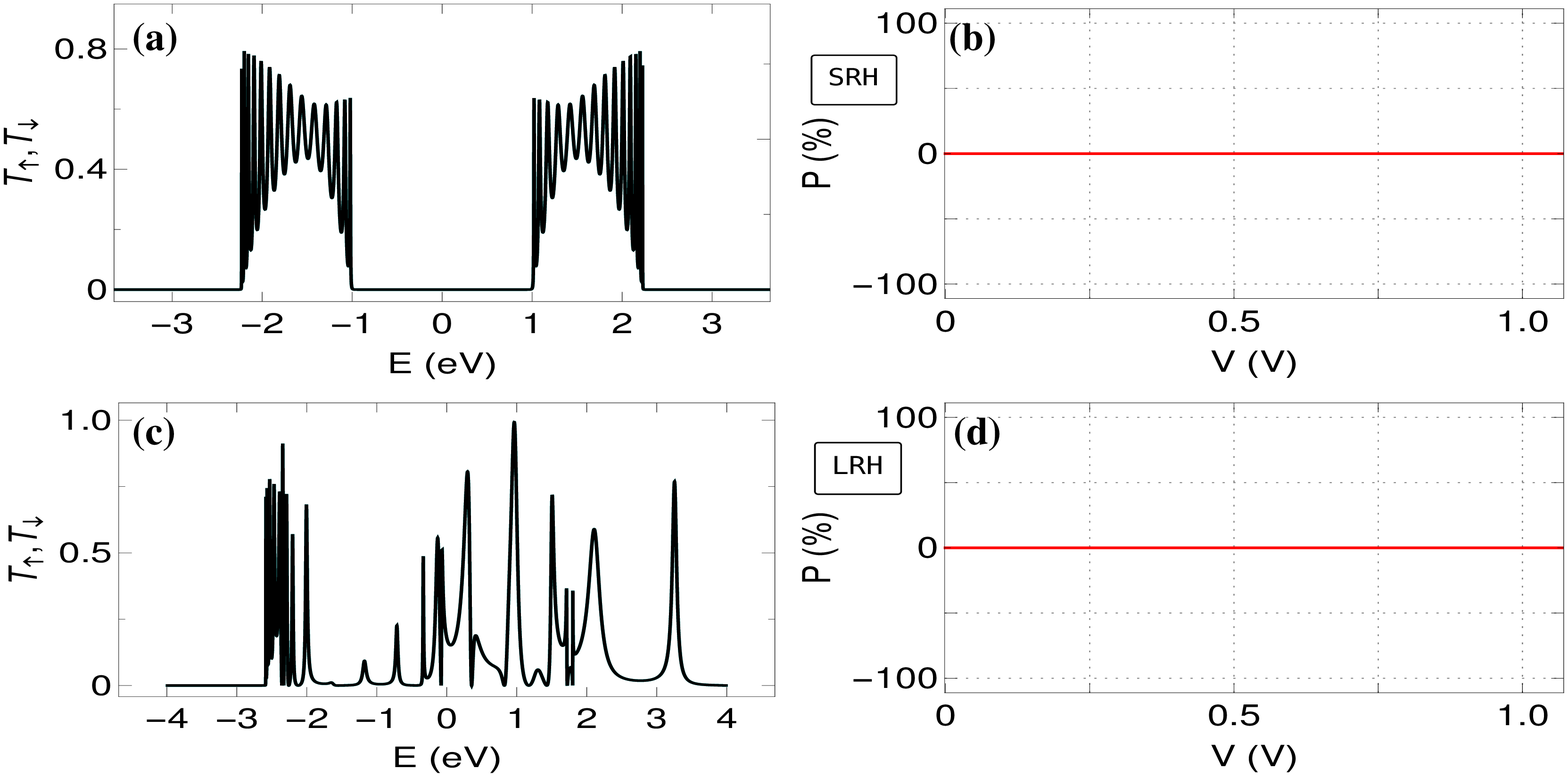}}\par}
\caption{(Color online). Zero field case: Spin dependent transmission probabilities as a function of energy for (a) SRH and (c) LRH AFHs,
where the black and cyan colors are associated with the up and down spin electrons, respectively. The up and down spin transmission 
spectra get exactly overlapped with each other for each AFH. The corresponding spin polarization coefficients with bias voltage are 
shown in (b)  and (d), respectively. Spin polarization drops to zero for the entire voltage window. Here we choose $E_F=-0.9\,$eV and 
set $N=30$. The dephasing strength $\eta=0$.}
\label{fig00}
\end{figure}

Let us begin with spin-dependent transmission probabilities, shown in the first column of Fig.~\ref{fig00}, where Fig.~\ref{fig00}(a)
and Fig.~\ref{fig00}(c) are associated with the SRH and LRH AFHs respectively. The corresponding spin polarization coefficients are
presented in the right column. All these results are evaluated for the electric field-free condition i.e. when $V_g = 0$. For both the
helix systems we find that up and down spin transmission probabilities, shown by the black and cyan colors respectively, exactly
overlap with each other, resulting in a vanishing spin polarization. This is expected, as for such helices the symmetry 
between the up and
down spin sub-Hamiltonians ($\boldsymbol{H}_{\uparrow}$ and $\boldsymbol{H}_{\downarrow}$) is preserved. Accordingly, we get an identical 
set of energy eigenvalues 
for the two different spin electrons, and thus the transmission probabilities, as the transmission peaks are directly related to the 
energy eigenvalues of the bridging system. The appearance of identical energy eigenvalues can easily be checked by writing the 
Hamiltonian of the AFH  ($\boldsymbol{\mathit{H}}_{AFH}$) as a sum of the two sub-Hamiltonians 
(viz, $\boldsymbol{\mathit{H}}_{AFH}=\boldsymbol{\mathit{H}}_{\uparrow} + \boldsymbol{\mathit{H}}_{\downarrow}$), one is associated 
with up spin electrons and the other is involved with the down spin ones. In the absence of the electric field, these two 
sub-Hamiltonians are effectively identical to each other, and hence, the same set of energy channels is obtained. For a particular AFH, 
the sharpness of different transmission peaks depends on the coupling ($t_S$ and $t_D$) of the helix to the side attached electrodes. 
With increasing the coupling, the transmission peaks get more broadened, and the broadening due to this coupling is always higher than 
that caused by the thermal effect. In our numerical calculation since $t_S$ and $t_D$ are comparable to $t_1$ (strong coupling limit),
any significant change with increasing temperature is not expected and therefore we restrict our calculation at a moderate temperature.

Depending on the specific range of electron hopping, we get a contrasting nature in the transmission peaks and their arrangements over
the energy window. For the chosen set of parameter values, electrons can able to hop in a few neighboring lattice sites in the SRH system,
and for this case, the transmission peaks are quite uniformly spaced and peaked as well. More regular behavior is obtained as we move
towards the NNH model. A finite and large gap is obtained across $E=0$, following the energy gap in the SRH system. This is quite 
analogous to the binary alloy system where alternate sites possess two different energies and repeat it throughout the system, due to the
antiferromagnetic ordering. But, once the higher-order hopping of electrons is taken into account, like what is considered for the LRH
system, the sharp gap around $E=0$ disappears. At the same time, the uniformity is lost significantly. Along one edge of the energy
window, the transmission peaks are closely packed whereas large gaps between the peaks are obtained along the other edge of the
window~\cite{protein}. All these characteristics are the generic features of a long-range hopping system. This asymmetric distribution,
on the other hand, plays an important role to achieve a favorable response in spin polarization, which can be visualized in the
forthcoming discussion.

\begin{figure}[ht]
{\centering\resizebox*{8cm}{6cm}{\includegraphics{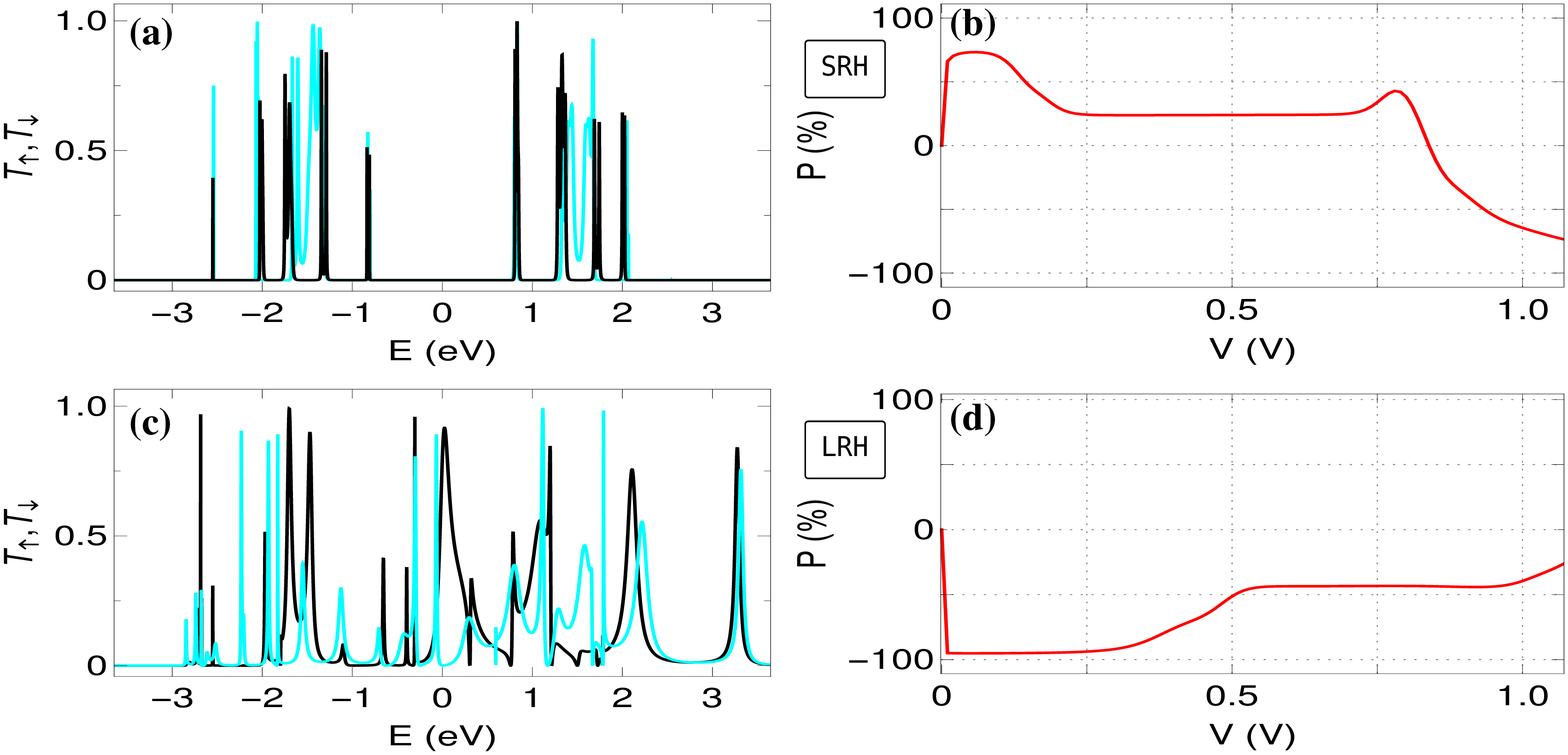}}\par}
\caption{(Color online). Results in the presence of electric field with $V_g=0.6\,$V. In (a) and (c), spin selective transmission 
probabilities as a function
of energy are shown for (a) SRH and (c) LRH AFHs, where the black and cyan colors are associated with up and down spin electrons,
respectively. The corresponding $P$-$V$ curves are shown in (b) and (d), respectively. All the other physical parameters are the same 
as taken in Fig.~\ref{fig00}.}
\label{fig01}
\end{figure}

The situation drastically changes, once we apply an electric field. In Fig.~\ref{fig01} we show the results, for the same set of systems
as taken in Fig.~\ref{fig00}, considering the gate voltage $V_g=0.6$\,V. Several notable features are obtained those are illustrated one
by one as follows. At a first glance, we find that up and down spin transmission probabilities get separated both for the SRH and LRH
helices, as clearly reflected from the spectra given in Figs.~\ref{fig01}(a) and (c), where two different colors are associated with two
different spin electrons. In the presence of electric field, site energies get modified in a cosine form, following Eq.~\ref{eg},
which makes the system a disordered (correlated) one~\cite{aah1,aah2,aah3}. Due to this disorder, the symmetry between up and down spin
sub-Hamiltonians gets broken, which provides different sets of spin-specific energy channels. Under this situation, a mismatch occurs 
between the spin-dependent transmission spectra. Apparently, it seems that the separation between the up and down spin transmission 
peaks is not that much large, what we generally expect from ferromagnetic systems, but selectively placing the Fermi energy, we can 
have the possibility of getting a reasonably high degree of spin filtration. This is precisely shown in Figs.~\ref{fig01}(b) and (d). 
Both for 
the SRH and LRH AFHs, large spin polarization is obtained over a particular voltage range, but the response becomes more favorable 
for the case of LRH AFH. This is entirely due to the non-uniform distribution of the transmission peaks around the center of the spectrum. 
So, naturally, starting from an NNH AFH, we can expect a better response whenever we include an additional hopping of electrons, and 
we confirm it through our detailed numerical calculations. Moreover, it is pertinent to note that, the cosine modulation in site 
energies due to the electric field makes the system a non-trivial one, as it generates a fragmented energy spectrum (which is the 
generic feature of the well-known Aubry-Andr\'{e}-Harper (AAH) model~\cite{aah1,aah2,aah3}). This behavior helps us to find a high 
degree of spin polarization even at multiple Fermi energies.

The key conclusion that is drawn from the above analysis is that the breaking of the symmetry between up and down spin sub-Hamiltonians 
in the AFH entirely depends on the external electric field which makes the system a disordered (correlated) one. In the absence of 
helicity, the site energies become uniform. Under this situation, the symmetry between the sub-Hamiltonians associated with up and down
spin electrons gets preserved, and hence, no such spin filtration phenomenon is obtained.

\subsection{Possible tuning of spin polarization}

This sub-section deals with the possible tuning of spin filtration efficiency, by adjusting different parameter values associated with
the junction setup.
\begin{figure}[ht]
{\centering\resizebox*{8cm}{6cm}{\includegraphics{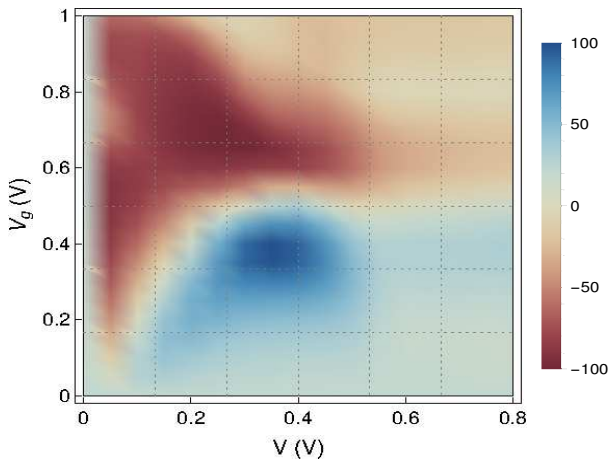}}\par}
\caption{(Color online). Simultaneous variations of $P$ with bias voltage $V$ and gate voltage $V_g$ for a $50$-site LRH AFH when $E_F$
is fixed at $1\,$eV. The dephasing parameter $\eta=0$.}
\label{fig1}
\end{figure}
From the above analysis since it is already established that LRH AFH is superior to the SRH AFH, in the rest part of our discussion
we concentrate only on the LRH AFH systems unless stated otherwise.

Figure~\ref{fig1} shows simultaneous variations of spin polarization coefficient $P$ with the bias voltage $V$ and the gate
voltage $V_g$. Both these two factors play a significant role in spin filtration efficiency. When $V_g$ is absolutely zero (i.e.,
in the absence of electric field) there is no spin polarization as up and down spin sub-Hamiltonians are {\em symmetric}.
With increasing $V_g$, we are expecting a favorable spin polarization, but attention has to be given to the localizing behavior of
energy eigenstates. The inclusion of an electric field transforms the AFH from a completely perfect to a correlated disordered system,
and thus, for large $V_g$ the eigenstates will be almost localized. In that limit, we cannot get spin filtration operation. Hence,
we need to restrict $V_g$ in such a way that the energy channels are conducting in behavior (see Ref.~\cite{sarkar} for a 
comprehensive
analysis of electronic localization in SRH and LRH helices in presence of transverse electric field). For a fixed $V_g$, when a finite
voltage drop is introduced across the junction, we get a non-zero spin polarization depending on the dominating energy channels among
up and down spin electrons. With the increase of the bias window, more and more number of both up and down spin channels are available
that contribute to the current, and hence, the possibility of mutual cancellations becomes higher which can reduce the degree of spin
polarization. Thus, both $V_g$ and the bias voltage need to be considered selectively, to have a favorable response.

Like bias voltage $V$, the choice of $E_F$ is also very crucial. Our aim is to find a suitable $E_F$ where any of the two spin channels 
dominates over the other, as maximum as it is possible. This, on the other hand, is directly linked with the gate voltage $V_g$ as it 
\begin{figure}[ht]
{\centering\resizebox*{8cm}{6cm}{\includegraphics{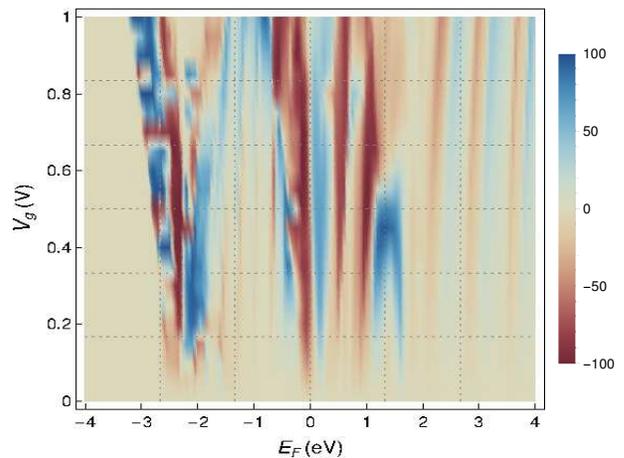}}\par}
\caption{(Color online). Dependence of $P$ with $E_F$ and $V_g$ when the bias voltage is fixed at $0.2\,$V. The other parameters are:
$N=50$ and $\eta=0$.}
\label{fig2}
\end{figure}
influences the site energies of the AFH. To have an idea about the selection of $V_g$ and $E_F$, in Fig.~\ref{fig2} we show the 
dependence of $P$ on these quantities. The results are computed, setting the voltage drop $V=0.2\,$V. This typical bias voltage is 
considered due to the fact that here we can get a favorable response as reflected from Fig.~\ref{fig1}. We vary $E_F$ almost over the 
entire allowed energy window, and it is seen that for a wide range of $E_F$ ($\sim -2 \leq E_F \leq \sim 2$), a reasonably large $P$ 
is obtained. Thus, fine tuning of $E_F$ is no longer required, which is of course quite important in the context of experimental 
realization of our proposed setup.

\begin{figure}[ht]
{\centering\resizebox*{8cm}{6cm}{\includegraphics{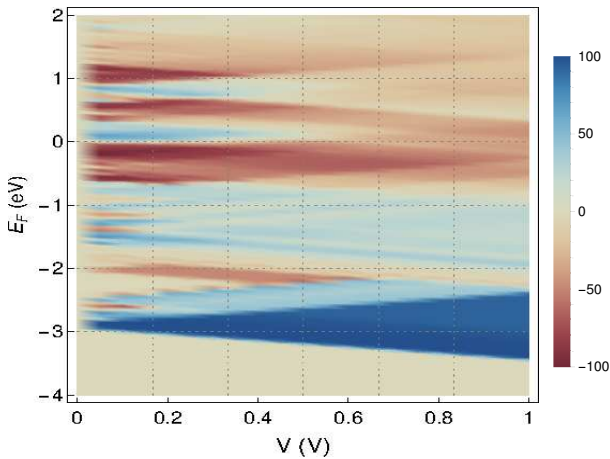}}\par}
\caption{(Color online). Dependence of $P$ with $E_F$ and $V$ when the gate voltage is fixed at $0.8\,$V. The other parameters are:
$N=50$ and $\eta=0$.}
\label{fig3}
\end{figure}

In the same footing it is indeed required to check the filtration efficiency when we simultaneously vary $E_F$ and the bias voltage $V$,
keeping the gate voltage constant. The results are shown in Fig.~\ref{fig3}, where we fix $V_g=0.8\,$V. This typical value of $V_g$ is
chosen observing the favorable response from Figs.~\ref{fig1} and \ref{fig2}. Here also we find that a high degree of spin polarization 
is achieved at different bias drops across the junction for several distinct choices of Fermi energy $E_F$. All these favorable responses
are associated with the modifications of the up and down spin energy channels.

Along with the favorable spin polarization, a complete phase reversal (change of sign) of $P$ is also obtained from all these figures 
(Figs.~\ref{fig1}-\ref{fig3}) which is due to the swapping of dominating spin channels with the change of the physical parameters.

\begin{figure}[ht]
{\centering\resizebox*{7cm}{4.5cm}{\includegraphics{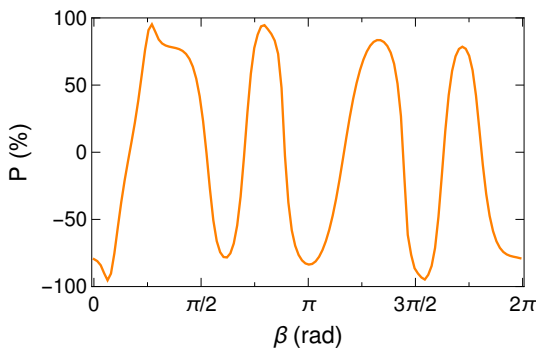}}\par}
\caption{(Color online). Variation of spin polarization ($P$) with the field direction ($\beta$). The other physical parameters are:
$E_F=0.5\,$eV, $V=0.2\,$V, $V_g=0.8\,$V, $N=50$ and $\eta=0$.}
\label{fig4}
\end{figure}

Now we concentrate on the effect of field direction, which is changed by the parameter $\beta$. The dependence of spin polarization with 
$\beta$ is presented in Fig.~\ref{fig4}, by varying $\beta$ from $0$ to $2\pi$. Almost a regular oscillation is shown providing high peaks 
($P\sim 100\%$) and dips ($P\sim -100\%$). The magnitude and sign reversal of $P$ are due to the modifications of spin specific energy 
channels of the helix with $\beta$ as it is directly related to the site energy (see Eq.~\ref{eg}). It suggests that the spin filtration 
efficiency can be monitored selectively by changing the field direction, keeping all the other factors unchanged.

\begin{figure}[ht]
{\centering\resizebox*{8cm}{6cm}{\includegraphics{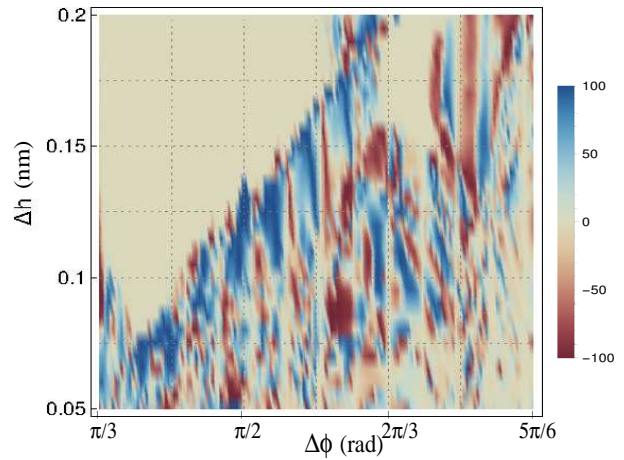}}\par}
\caption{(Color online). Simultaneous variations of $P$ with $\Delta \phi$ and $\Delta h$. Here we choose $E_F=0.5\,$eV, $V=0.2\,$V, 
$V_g=0.8\,$V, $N=50$, $\beta=0$ and $\eta=0$.}
\label{fig5}
\end{figure}

Following the above analysis it is found that the helicity and electric field are strongly correlated with each other, and in the absence 
of any of these two, spin filtration is no longer possible. At this stage, it is indeed required to check the effect of helicity on a more
deeper level.
In Fig.~\ref{fig5} we show the conformational effect of the helical geometry on spin polarization. We vary $\Delta h$ and $\Delta \phi$ 
in a reasonable range around the chosen values of these parameters for the LRH AFH, as mentioned in Table~\ref{table1}. The radius $R$ 
is kept constant which is $0.25\,$nm. The twisting angle $\Delta \phi$ has an important role as it modulates the site energy as well as 
the hopping integrals. But the more pronounced effect is observed by changing the stacking distance $\Delta h$. With the enhancement of 
$\Delta h$, for a fixed $\Delta \phi$, the hopping of electrons in larger sites gradually decreases, and the system approaches the NNH 
model, providing reduced spin polarization. In the limiting region when electrons can hop only between the nearest-neighbor 
sites, the spin polarization becomes vanishingly small. Carefully inspecting the density plot given in Fig.~\ref{fig5}, it is 
inferred that the best performance is obtained for low $\Delta h$ when the twisting angle is confined within the range 
$\sim \pi/2 \le \phi \le \sim 2\pi/3$. Thus, the helicity and higher-order hopping of electrons are the key aspects to having a high 
degree of spin filtration.

\subsection{Effect of dephasing}

To have a more realistic situation, especially considering the experimental facts, in this sub-section we explore the specific role of 
dephasing~\cite{dephase1,dephase2,dephase3,dephase4,dephase5} that may enter into the system in many ways on spin polarization. Different 
kinds of interactions of the physical system with external factors can phenomenologically be incorporated through the dephasing effect.

\begin{figure}[ht]
{\centering\resizebox*{6cm}{8cm}{\includegraphics{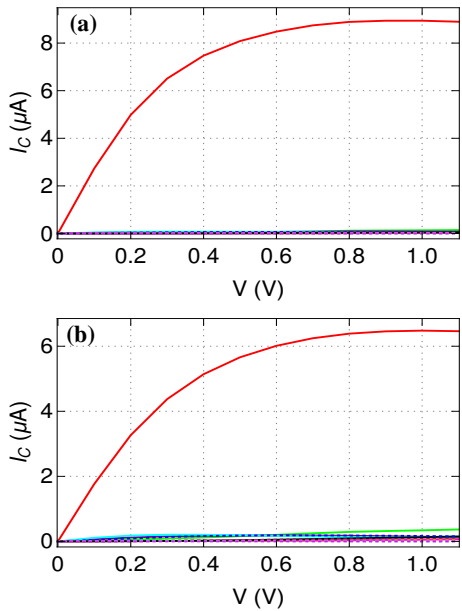}}\par}
\caption{(Color online). Current in the drain electrode (red curve) along with the currents in all the B\"{u}ttiker probes (other 
colored curves, not clearly seen as they almost overlap to each other) as a function of bias voltage for a spin-less LRH helix,
where (a) $\eta=0.25$ and (b) $\eta=0.5$. The other physical parameters are: $N=15$, $E_F=0.3\,$eV and $V_g=0$.}
\label{bp}
\end{figure}

\vskip 0.2cm
\noindent
$\blacksquare$ {\bf Accuracy checking of theoretical prescription in presence of dephasing}: 
Before presenting the results in the presence of dephasing, it is indeed required to check the accuracy of the theoretical formulation
illustrated above in Sec. II. To do that, we consider a spin-less LRH helix with $15$ lattice sites, as a typical example, and compute
the currents in all the B\"{u}ttiker probes (such probes are connected in all the sites of the helix from the site number $2$ to $14$),
along with the drain current (red curve). The results are shown in Fig.~\ref{bp} for the two distinct dephasing strengths that are 
presented in (a) and (b), setting the Fermi energy $E_F=0.3\,$eV. It is clearly seen that the drain current is reasonably large than 
the currents in all the B\"{u}ttiker probes, and most interestingly we find that the currents in the B\"{u}ttiker probes are almost zero,
even for too high voltages. This is exactly what we are expecting i.e., the vanishing current condition in each dephasing 
electrode. Thus, we can argue that the theoretical prescription given here can safely be used to study the effect of dephasing.

\begin{figure}[ht]
{\centering\resizebox*{8cm}{6cm}{\includegraphics{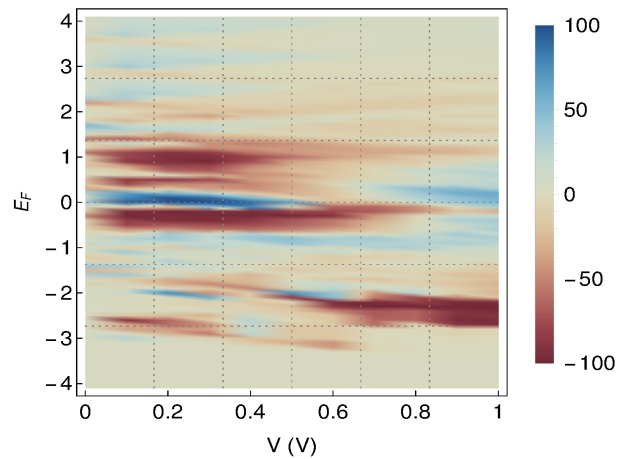}}\par}
\caption{(Color online). A similar kind of density plot, as given in Fig.~\ref{fig3}, in the presence of dephasing, with the dephasing
strength $\eta = 0.1$.}
\label{fig8}
\end{figure}

Now come to the results of AFH in presence of dephasing. 
Like Fig.~\ref{fig3}, in Fig.~\ref{fig8} we show the simultaneous variations of $P$ on the bias voltage $V$ and the Fermi energy $E_F$,
fixing the dephasing strength $\eta=0.1$. All the other physical parameters are kept unchanged as taken in Fig.~\ref{fig3}. The effect of 
dephasing is quite appreciable. What we see is that the Fermi energy and the bias windows for which a high degree of spin polarization is
available in the absence of dephasing (see Fig.~\ref{fig3}), get reduced when the dephasing effect is taken into account. The reduction
of spin polarization in presence of $\eta$ can be explained as follows. In the B\"{u}ttiker probe prescription, the effect of dephasing 
is incorporated by connecting each and every lattice site of the AFH with virtual electrodes. Due to the coupling of the AFH to these 
virtual electrodes, transmission peaks get broadened, and thus more overlap takes place between the up and down spin channels. Therefore,
these two spin-dependent channels contribute to the current, and hence, the spin polarization decreases. A similar kind of dephasing 
effect (viz, reduction of $P$ with $\eta$) is also obtained when we observe simultaneous variations of $P$ with $V_g$ and $V$ keeping 
$E_F$ constant, and, $V_g$ and $E_F$ considering a fixed bias voltage. Accordingly, here we do not present the density plots of $P$, 
like what are shown in Fig.~\ref{fig1} and Fig.~\ref{fig2}, in the presence of $\eta$ as the role $\eta$ can be guessed in these cases.

\begin{figure}[ht]
{\centering\resizebox*{6cm}{6.5cm}{\includegraphics{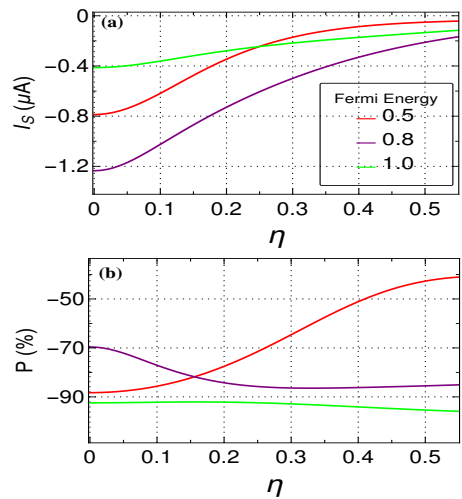}}\par}
\caption{(Color online). (a) Spin current and (b) spin polarization coefficient as a function of dephasing strength $\eta$, at three
typical Fermi energies. Here we set $N=50$, $V=0.2\,$V and $V_g=0.8\,$V.}
\label{fig7}
\end{figure}

Here it is relevant to check the dependence of spin current ($I_S=I_{\uparrow}-I_{\downarrow}$) and spin polarization for other values of
the dephasing strength $\eta$ as well. The results are given in Figs.~\ref{fig7} (a) and (b), respectively, by varying $\eta$ in a 
reasonable range. Three 
different Fermi energies are taken into account, those are represented by three different colored curves. It is clearly observed that 
the spin current decreases with increasing $\eta$, following the above argument of the broadening of both up and down spin transmission 
peaks and their overlap (Fig.~\ref{fig7}(a)). This reduction of spin current can also be implemented by means of phase randomization of 
the electrons in the virtual electrodes, as originally put forward by B\"{u}ttiker~\cite{dephase1,dephase2,dephase5}. Electrons enter 
into these electrodes and they come back
to the helix system after randomizing their phases, and thus more mixing of the two opposite spin electrons occurs, resulting in a 
reduced spin current. In the same analogy, the degree of spin polarization gets decreased with increasing the dephasing strength 
(Fig.~\ref{fig7}(b)). From $P$-$\eta$ characteristics it is found that though the spin filtration efficiency gets reduced with $\eta$,
still a sufficiently large value of $P$ is obtained even when $\eta$ is reasonably large. Thus, we can safely claim that our proposed 
quantum system can be utilized as an efficient functional element for spin filtration operation under strong environmental interactions
as well as the limit of high temperatures.

\subsection{Possible experimental routes of designing AFH}

Finally, we refer to some experimental works where different kinds of anti-ferromagnetic systems have been used, 
aiming to establish confidence that our proposed magnetic helix system can also be realized experimentally. In presence of an external 
magnetic field, Johnston has reported~\cite {afhelix1} magnetic properties 
and other related phenomena considering an AFH system. There are several experimental works performed by Sangeetha and co-workers, 
where they have found helical antiferromagnet in different compounds. For instance in Ref.~\cite{afhelix2}, Sangeetha {\em et al.} have 
established a transition from antiferromagnetic to paramagnetic phase using the compound ${\rm Eu Co}_{2-y}{\rm As}_2$ with spin 
$S=7/2$ which possesses a helical shape~\cite{afhelix2}. In this work, they have measured different physical quantities like magnetic 
susceptibility, heat capacities, etc. A high nuclear magneto resistance (NMR) has also been found in that sample~\cite{nmr}. In another 
work, Sangeetha and co-workers have established helical antiferromagnetic ordering considering ${\rm Eu Ni}_{1.95} {\rm As}_2$ single 
crystal~\cite{afhelix3}. Goetsch {\em et al.} have reported the same in polycrystalline sample ${\rm Lu}_{1-x} {\rm Sc}_x {\rm MnSi}$ at 
higher Ne\'el's temperature~\cite{afhelix4}. The antiferromagnetic helical pattern has also been noticed in an organic molecule. 
Lin {\em et al.} have reported the canted antiferromagnetic behavior in ${\rm [M (mtpo)_2 (H_2O)_n]}$, ${\rm M = Co^{2+}}$ or 
${\rm Ni^+}$ with a helical topology~\cite{afhelix5}. Pylypovskyi {\em et al.} have suggested an idea about the tailoring of the geometry 
of curvilinear antiferromagnet~\cite{afhelix6}. This prescription allows substantiating a chiral helimagnet in presence of 
Dzyloshinskii-Moriya interaction. Considering the helical antiferromagnet sample ${\rm Sr Fe O}_{3-\delta}$, Zhao {\em et al.} have 
discussed a metal-insulator transition~\cite{afhelix7}. There exist several other antiferromagnetic helices as well.

Considering all such examples of antiferromagnetic helical systems, we believe that our proposed spin-polarized AFH system can be designed 
with modern technology and with a suitable laboratory setup. 

Here it is relevant to note that all the above-mentioned experimental references contain heavy magnetic elements, and 
thus one may think whether the tight-binding Hamiltonian mentioned in Eq.~\ref{hamil} can be used to describe our helix systems or not. 
But the theoretical work studied by Takahashi and Igarashi~\cite{taka} gives us confidence that Eq.~\ref{hamil} can safely be considered, 
as in that work they have also taken a similar kind of tight-binding Hamiltonian to describe ${\rm La_2CuO_4}$ and ${\rm Sr_2CuO_2Cl_2}$.
There exist several other references as well~\cite{pic,cao} where tight-binding Hamiltonians have been taken into account for such 
types of heavy magnetic elements.

\section{Summary and outlook}

For the first time, we report spin filtration operation considering an antiferromagnetic helical system in presence of an external
electric field. Both the short-range and long-range hopping cases are taken into account, associated with the geometrical conformation.
Simulating the spin-polarized nanojunction (source-AFH-drain) within a tight-binding framework, we compute spin-dependent transmission
probabilities following the well-known Green's function formalism and the spin-dependent junction currents through the
Landauer-B\"{u}ttiker prescription. From the currents, we evaluate the spin polarization coefficient. To make the proposed quantum system
more realistic, we also include the effects of dephasing, and to get the confidence, the accuracy of the theoretical 
prescription in presence of dephasing is critically checked. Finally, we discuss the possible routes of realizing such an 
antiferromagnetic helical geometry in a laboratory. Different aspects of spin-dependent transmission probabilities and spin polarization 
coefficient under different input conditions are critically investigated. The essential findings are listed as follows. \\
$\bullet$ In the absence of the external electric field, up and down spin sub-Hamiltonians become symmetric and thus no spin 
polarization is obtained. Once the symmetry is broken by applying the electric field, finite spin polarization is found. \\
$\bullet$ Comparing the results between the SRH and LRH AFHs, it is noticed that LRH AFH is much superior to the other. This is
essentially due to the irregular distribution of the resonant peaks in the transmission-energy spectrum. The irregularity gradually
decreases with lowering the electron hopping among the lattice sites. \\
$\bullet$ The degree of spin polarization and its phase can be tuned selectively by means of the input parameters, and the notable
thing is that a high degree of spin polarization persists over a reasonable range of physical parameters. It clearly suggests that fine
tuning of the parameter values is no longer required, and hence we hope that the studied results can be examined in a laboratory. \\
$\bullet$ The geometrical conformation plays an important role. For the situation when $\Delta \phi$ becomes zero i.e., in the absence
of twisting, spin filtration is no longer available. \\
$\bullet$ Though the degree of spin filtration efficiency gets reduced with increasing the dephasing strength $\eta$, still, reasonably
large spin polarization is available, even for moderate $\eta$. It suggests that the proposed functional element can safely be used for
spin polarization under strong environmental interactions as well as in the limit of high temperatures.

Our present proposition may help to design efficient spintronic devices using antiferromagnetic helices with longer-range hopping of
electrons, and can be generalized to other correlated antiferromagnetic systems as well.

\end{document}